\documentclass[twocolumn,aps,superscriptaddress,prl]{revtex4}%%,twocolumn
\usepackage{graphicx}
\usepackage{xcolor}
\usepackage{times}
\usepackage{amsmath}
\usepackage{amssymb}
\usepackage{units}
\usepackage{stmaryrd} %% "// sign" = \sslash
\DeclareGraphicsExtensions{.pdf,.png,.eps,.jpg}

\begin{document}
\preprint{0}

\title{Enhanced ultrafast relaxation rate in the Weyl semimetal  phase of $\mathbf{MoTe_2}$ measured by time-and angle-resolved photoelectron spectroscopy}

\author{A. Crepaldi} 
\affiliation{Institute of Physics, Ecole Polytechnique F\'ed\'erale de Lausanne (EPFL), CH-1015 Lausanne, Switzerland} 

\author{G. Aut\`es}
\affiliation{Institute of Physics, Ecole Polytechnique F\'ed\'erale de Lausanne (EPFL), CH-1015 Lausanne, Switzerland}
\affiliation{National Centre for Computational Design and Discovery of Novel Materials MARVEL, Ecole Polytechnique F\'ed\'erale de Lausanne (EPFL), CH-1015 Lausanne, Switzerland}

\author{G. Gatti} 
\affiliation{Institute of Physics, Ecole Polytechnique F\'ed\'erale de Lausanne (EPFL), CH-1015 Lausanne, Switzerland} 

\author{S. Roth} 
\affiliation{Institute of Physics, Ecole Polytechnique F\'ed\'erale de Lausanne (EPFL), CH-1015 Lausanne, Switzerland} 

\author{A. Sterzi}
\affiliation{Universit\'a degli Studi di Trieste - Via A. Valerio 2, Trieste 34127, Italy} 

\author{G. Manzoni} 
\affiliation{Universit\'a degli Studi di Trieste - Via A. Valerio 2, Trieste 34127, Italy} 

\author{M. Zacchigna}
\affiliation{C.N.R. - I.O.M., Strada Statale 14, km 163.5, Trieste 34149, Italy}

\author{C. Cacho}
\affiliation{Central Laser Facility, STFC Rutherford Appleton Laboratory, Harwell OX11 0QX, United Kingdom}

\author{R. T. Chapman}
\affiliation{Central Laser Facility, STFC Rutherford Appleton Laboratory, Harwell OX11 0QX, United Kingdom}

\author{E. Springate}
\affiliation{Central Laser Facility, STFC Rutherford Appleton Laboratory, Harwell OX11 0QX, United Kingdom}

\author{E. A. Seddon}
\affiliation{The Photon Science Institute, The University of Manchester, Manchester, United Kingdom }
\affiliation{The Cockcroft Institute, Sci-Tech Daresbury, Daresbury, Warrington, United Kingdom }

\author{Ph. Bugnon}
\affiliation{Institute of Physics, Ecole Polytechnique F\'ed\'erale de Lausanne (EPFL), CH-1015 Lausanne, Switzerland}

\author{A. Magrez}
\affiliation{Institute of Physics, Ecole Polytechnique F\'ed\'erale de Lausanne (EPFL), CH-1015 Lausanne, Switzerland}

\author{H. Berger}
\affiliation{Institute of Physics, Ecole Polytechnique F\'ed\'erale de Lausanne (EPFL), CH-1015 Lausanne, Switzerland}

\author{I. Vobornik}
\affiliation{CNR. - IOM., Strada Statale 14, km 163.5, Trieste 34149, Italy}

\author{M. Kall\"ane}
\affiliation{Institut f\"ur Experimentelle und Angewandte Physik, Christian-Albrechts-Universit\"at zu Kiel, D-24098 Kiel, Germany}

\author{A. Quer}
\affiliation{Institut f\"ur Experimentelle und Angewandte Physik, Christian-Albrechts-Universit\"at zu Kiel, D-24098 Kiel, Germany}

\author{K. Rossnagel}
\affiliation{Institut f\"ur Experimentelle und Angewandte Physik, Christian-Albrechts-Universit\"at zu Kiel, D-24098 Kiel, Germany}

\author{F. Parmigiani}
\affiliation{Universit\'a degli Studi di Trieste - Via A. Valerio 2, Trieste 34127, Italy} 
\affiliation{Elettra - Sincrotrone Trieste  S.C.p.A., Strada Statale 14, km 163.5, Trieste 34149, Italy} 
\affiliation{International Faculty - University of K\"oln, 50937 K\"oln, Germany} 

\author{O. V. Yazyev}
\affiliation{Institute of Physics, Ecole Polytechnique F\'ed\'erale de Lausanne (EPFL), CH-1015 Lausanne, Switzerland}
\affiliation{National Centre for Computational Design and Discovery of Novel Materials MARVEL, Ecole Polytechnique F\'ed\'erale de Lausanne (EPFL), CH-1015 Lausanne, Switzerland}

\author{M. Grioni} 
\affiliation{Institute of Physics, Ecole Polytechnique F\'ed\'erale de Lausanne (EPFL), CH-1015 Lausanne, Switzerland}

\begin{abstract}

$\mathrm{MoTe_2}$ has recently been shown to realize in its low-temperature phase the type-II Weyl semimetal (WSM). We investigated by time- and angle- resolved photoelectron spectroscopy (tr-ARPES) the possible influence of the Weyl points in the electron dynamics above the Fermi level $\mathrm{E_F}$, by comparing the ultrafast response of $\mathrm{MoTe_2}$ in the trivial and topological phases. In the low-temperature WSM phase, we report an enhanced relaxation rate of electrons optically excited to the conduction band, which we interpret as a fingerprint of the local gap closure when Weyl points form. By contrast, we find that the electron dynamics of the related compound $\mathrm{WTe_2}$ is slower and temperature-independent, consistent with a topologically trivial nature of this material. Our results shows that tr-ARPES is sensitive to the small modifications of the unoccupied band structure accompanying the structural and topological phase transition of $\mathrm{MoTe_2}$.

%\textcolor{blue} {The dispersion of topological Fermi arcs have been reported by angle-resolved photoelectron spectroscopy (ARPES), however the physics of the Weyl points eludes direct detection, since the points are located above the Fermi level $\mathrm{E_F}$.} 

\end{abstract}

\date{\today}

\maketitle

%------------------------------------------------------------

%----------------

%------------------------------------------------------------	

The recent discovery of Weyl fermions as low-energy quasiparticles in TaAs \cite{Xu_Science_2015, Yang_natPhys_2015, Lv_NatPhys_2015} and other related compounds \cite{Xu_NatComm_2016, Autes_PRL_16} has boosted the interest in topological semimetals (TSMs) \cite{Burkov_NatMat_16}. Type-II Weyl semimetals (WSMs) are a novel class of materials that host fermions violating Lorentz invariance \cite{Soluyanov_Nature_2015}. These quasiparticles are realized in the strongly tilted cones that form in momentum space around special Weyl points (WPs) where the valence and conduction bands touch. $\mathrm{WTe_2}$ \cite{Soluyanov_Nature_2015} and $\mathrm{MoTe_2}$ \cite{Sun_PRB_2015, Wang_PRL_2016, Chang_NatCom_16} have been proposed as possible type-II WSMs.  While the topological phase of $\mathrm{WTe_2}$ is still under debate, the existence of the WSM phase in the low temperature non-centrosymmetric structure of $\mathrm{MoTe_2}$ is supported by the observation of surface Fermi arcs \cite{Tamai_PRX_16, Huang_NatMat_2016, Deng_NatPhy_2016, Jiang_NatCom_2016, Sakano_PRB_17}.  The Weyl points, however, are located above the Fermi level $\mathrm{E_F}$ and this hinders a direct observation by conventional angle-resolved photoelectron spectroscopy (ARPES). To circumvent this difficulty, one would need a probe that is sensitive to the presence/absence of small energy band gaps in the unoccupied density of states. 

\begin{figure*}[t!]
   \includegraphics[width = 0.75 \textwidth]{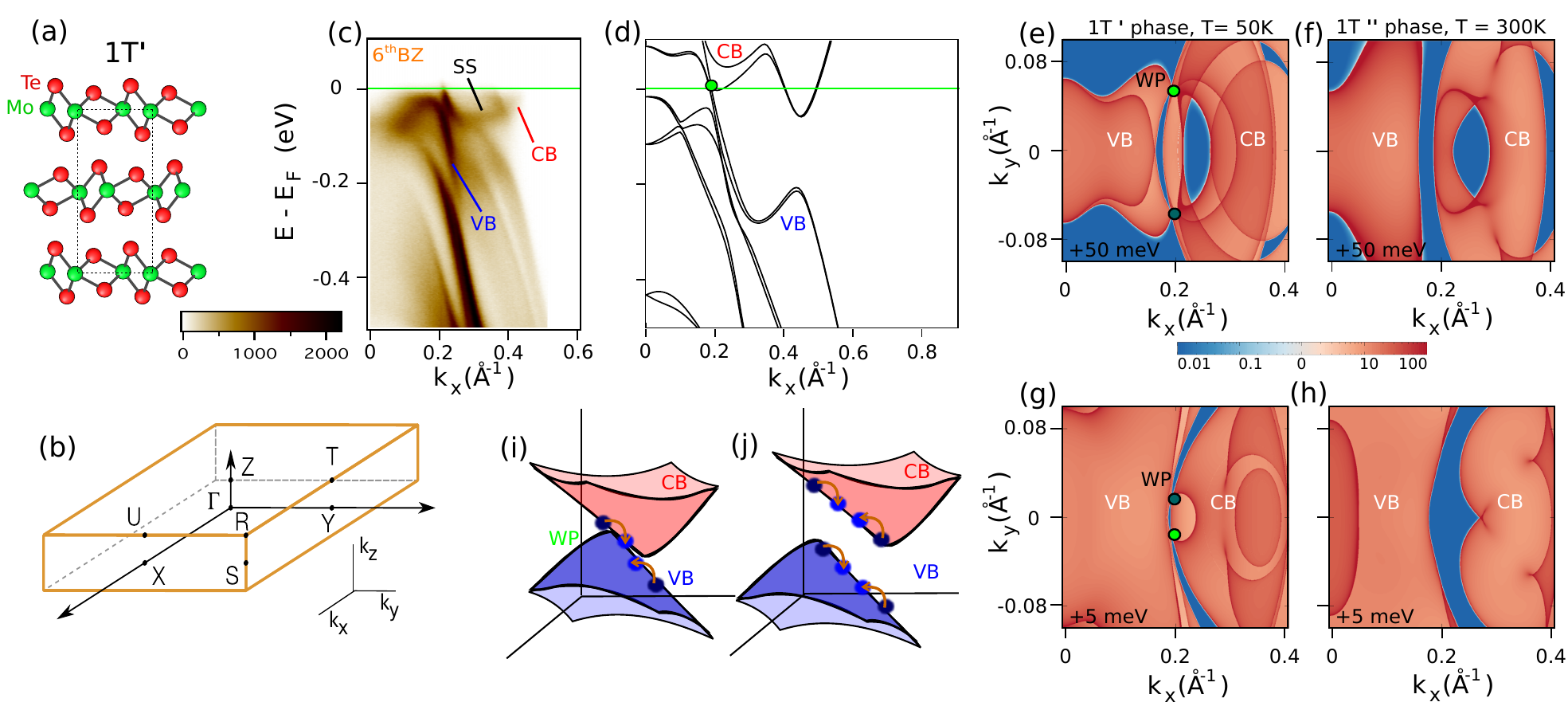}
  \caption{Crystal structure (a) and three-dimensional Brillouin zone (b) of 1T' $\mathrm{MoTe_2}$ \cite{Hughes_1978}. (c, d)  Experimental and calculated band dispersion of 1T' $\mathrm{MoTe_2}$ along the $\mathrm{\Gamma X}$ high-symmetry direction.  (e - h) Calculated $k$-resolved bulk density of states, projected on the [001] surface BZ, for the low- (e, g) and high- (f, h) temperature phases of $\mathrm{MoTe_2}$, at two different binding energies.  The presence of two pairs of WPs is revealed in the 1T' phase, at 50 meV (e) and 5 meV (g), respectively. In the centrosymmetric 1T'' phase,  the WPs vanish and the VB and the CB are well separated in momentum and in energy, as shown in the the CEMs at 50 meV  (f) and 5 meV (h). (i, j) The cartoon shows that electron - electron interaction can mediate the scattering between CB and VB along the Weyl cone in the WSM phase (i). Interband scattering requires momentum and energy exchange in the presence of the gap (j).}

 \label{fig:ARPES_MoTe}
 
 \end{figure*}

In this Letter we show that time- and angle-resolved photoelectron spectroscopy (tr-ARPES) can provide such information. We report a shortening of the relaxation time in the gapless type-II Weyl phase of $\mathrm{MoTe_2}$, which reflects the enhanced interband scattering from the conduction band (CB) to the valence band (VB) mediated by electron -- electron scattering along the Weyl cone.  These scattering processes are  active only when the band gap is locally closed. This conclusion is supported by the observation of  a slower, temperature-independent dynamics in $\mathrm{WTe_2}$, indicative of a local direct band gap, which acts as an effective bottleneck for the electron relaxation.

$\mathrm{MoTe_2}$ is a layered material. It can be cleaved to expose large flat $(001)$ terraces, ideal for ARPES studies. Figure\,1 illustrates the WSM phase, which is only realized in the low-temperature orthorhombic (space group P$mn2_1$) structure -- hereafter referred to as the 1T' phase \cite{note} -- where inversion symmetry is broken. The crystal structure is sketched in Fig.\,1\,(a), along with the 3D Brillouin zone (BZ) (Fig.\,1\,(b)). Above $250$\,K the 1T' structure transforms to the centrosymmetric monoclinic (space group P$12/m1$) 1T'' structure \cite{Hughes_1978}
% -- referred to elsewhere as       $\beta$ phase \cite{Tamai_PRX_16} and  also as 1T' phase \cite{Huang_NatMat_2016, Deng_NatPhy_2016, Jiang_NatCom_2016}. 
Raman spectroscopy and resistivity measurements indicate a structural phase transition at T* $\sim 257.5$\,K \cite{Zhang_Raman}. Details about the crystals growth and their characterization can be found in the Supplementary Information \cite{SuppInfo}.

%-------------------

The ARPES experiments in the UV photon energy range have been performed at the APE beamline at Elettra, and soft x-ray ARPES experiments at the end station ASPHERE III of beamline P04 at PETRA III (DESY). The tr-ARPES experiments have been carried out at the ARTEMIS facility, using ultrafast pulses at $17.5$\,eV photon energy, produced by laser-induced high harmonics generation (HHG) in a gas. The optical excitation was driven by a $2$~eV pump pulse with fluence $\mathrm{\sim0.3\,m J / cm^{2} }$. The energy and temporal resolutions were $150$~meV and $50$~fs, respectively \cite{SuppInfo}. 

Figure 1\,(c) illustrates the band structure of the 1T' phase along the $\mathrm{\Gamma X}$ direction. The VB consists of several hole-like states, in good agreement with the literature  \cite{Huang_NatMat_2016,  Tamai_PRX_16, Deng_NatPhy_2016, Jiang_NatCom_2016, Crepaldi_PRB_17}. The bottom of the CB forms a shallow pocket around $\mathrm{k_x  =} \pm0.4$\, \AA $^{-1}$, Fig.\,1\,(c). A detailed mapping of the bulk band structure is shown in \cite{SuppInfo}.
Our \emph{ab initio} band structure calculations of 1T' $\mathrm{MoTe_2}$ predict four pairs of WPs of two different types where the VB and the CB touch at $\mathrm{k_x} \sim 0.2$\,\AA$^{-1}$, slightly off the $\mathrm{\sim\Gamma X}$ direction, respectively $5$\,meV and $50$\,meV above $\mathrm{E_F}$ (Fig.\,1\,(d)). The WPs are located near the local minimum of the CB, which suggests the possibility of electron accumulation. Their ($\mathrm{k_x, k_y}$) locations are illustrated in Fig.\,1\,(e, g). By contrast, a similar calculation yields no WP in the trivial high-temperature 1T'' phase, where a local gap opens between the VB and the CB, Fig.\,1\,(f, h) \cite{SuppInfo}.

%-------------------
%-------------------

\begin{figure}[b]
   \includegraphics[width = 0.5 \textwidth]{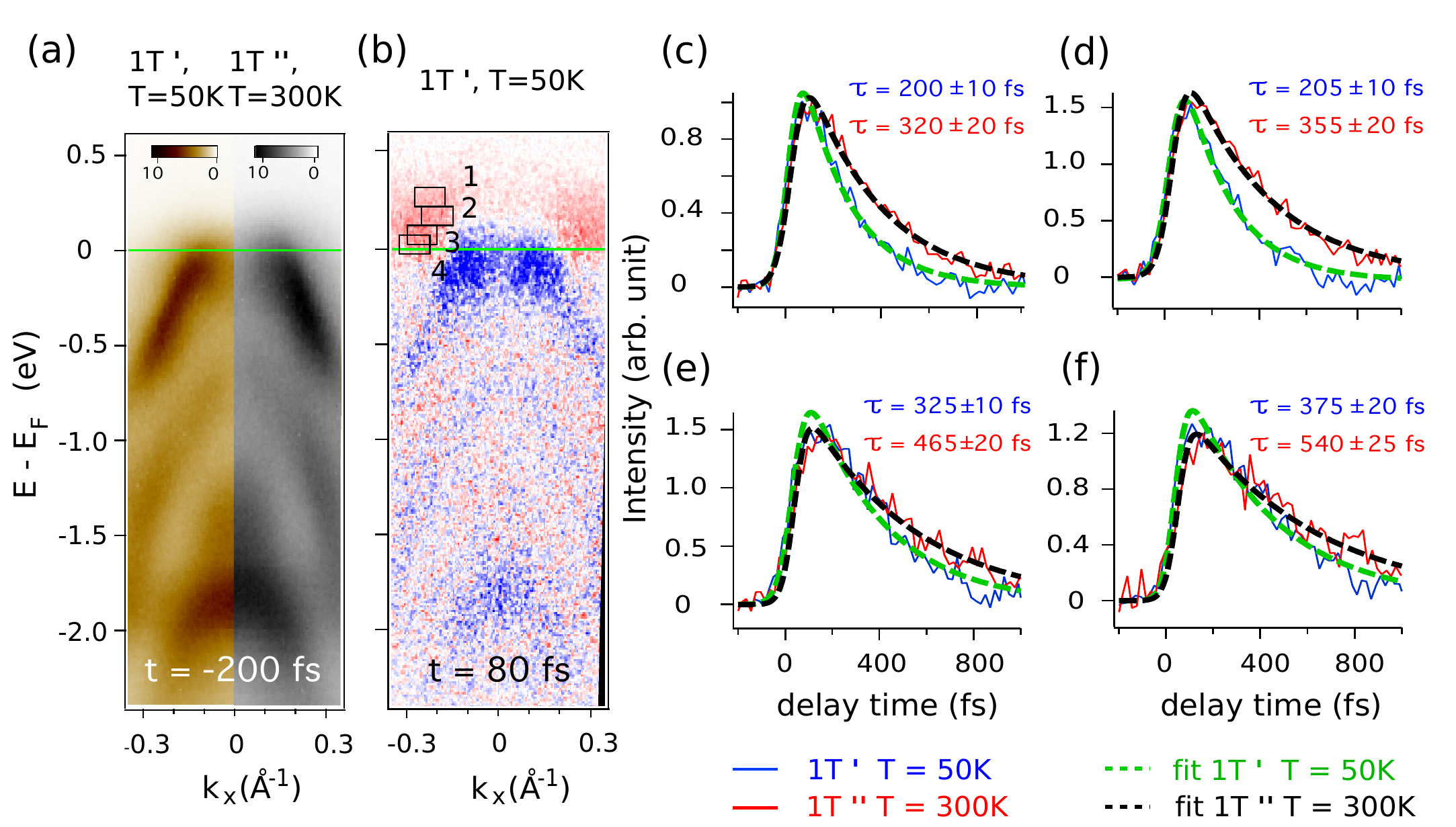} 
\caption[didascalia]{(a) Band dispersion of $\mathrm{MoTe_2}$ along $\mathrm{\Gamma X}$ at low- (left) and high- (right) temperatures, measured with 17.5 eV HHG photons $200$n fs before optical excitation. 
(b) Effect of optical excitation on 1T' $\mathrm{MoTe_2}$. Difference between  ARPES images acquired 80 fs after and 200 fs before the arrival of a $2$~eV optical excitation with fluence $\mathrm{\sim0.3\,m J/ cm^{2} }$. Red and blue indicate the increase and decrease in the electron population, respectively. (c - f) Dynamics in the regions 1 - 4 of panel (b) near the WPs for the 1T' (blue) and 1T'' (red) phases.  A  fit to the traces (dashed green and black lines) yields quantitative estimates of the characteristic relaxation times, $\mathrm{\tau}$.
 	 } 

\label{fig:tr_ARPES_MoTe}
 
 \end{figure}
 
%-------------------

%-------------------

\begin{figure*}[]
\includegraphics[width = 1.0 \textwidth]{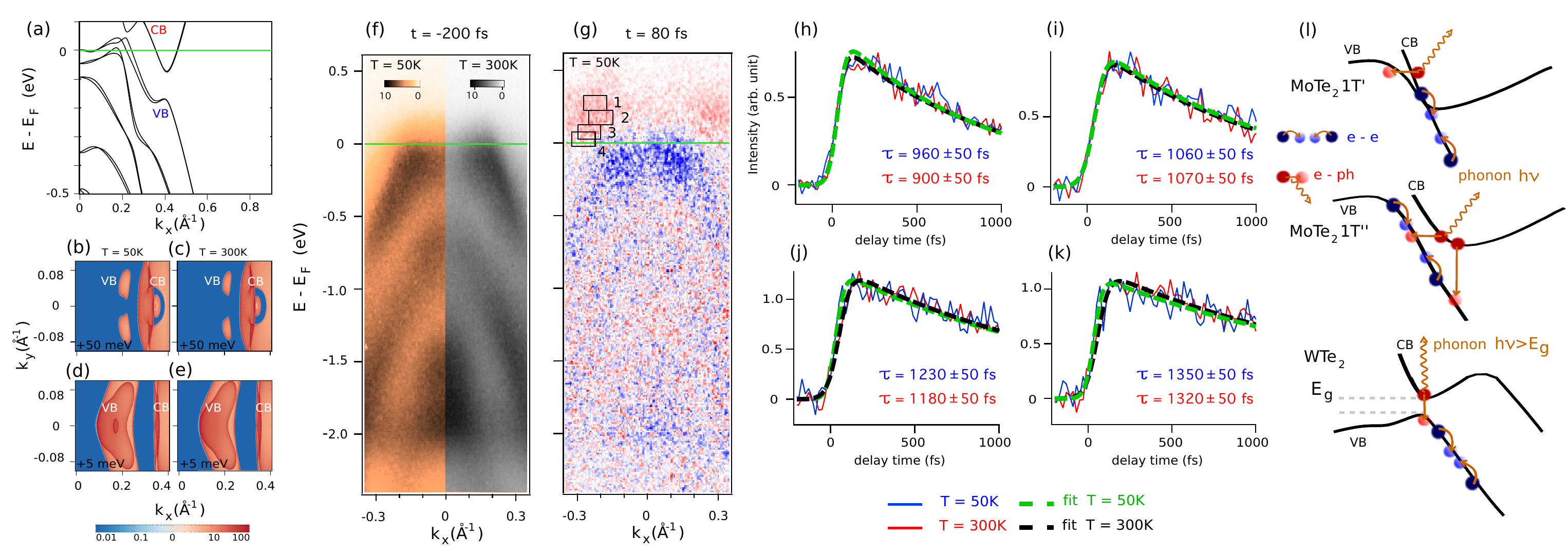}
\caption[didascalia]{(a) Band structure of $\mathrm{WTe_2}$ along $\mathrm{\Gamma X}$ calculated for the low-temperature crystal lattice parameters.  (b - e) Calculated $k$-resolved bulk density of states, projected on the [001] surface BZ, at two different binding energies for the crystal structure at low and high temperatures. Unlike the case of $\mathrm{MoTe_2}$ the CB and VB are always well separated in momentum and energy. 
(f) Band structure along $\mathrm{\Gamma X}$ for $\mathrm{WTe_2}$ measured at $50$~K (left) and at $300$~K (right) with $17.5$~eV HHG photons.  (g) The difference between  ARPES images acquired 80 fs after and 200 fs before the arrival of a $2$~eV optical excitation with fluence $\mathrm{\sim0.3\,m J/ cm^{2} }$. (h - k) Comparisons of the dynamics in  regions 1 - 4 at low (blue) and high (red) temperature.  A fit to the traces (dashed green and black lines) yields quantitative estimations of the characteristic relaxation time, $\mathrm{\tau}$, which is three times larger than the gapped 1T'' $\mathrm{MoTe_2}$. (l) Close-up on the calculated band structures of 1T'  $\mathrm{MoTe_2}$, 1T''  $\mathrm{MoTe_2}$ and  $\mathrm{WTe_2}$ in the region of the WPs. The possible electron -- electron and electron -- phonon scattering events are shown schematically. 
   	}

  \label{fig:tr_ARPES_WTe}
 
 \end{figure*}

%-------------------

Unlike conventional ARPES, tr-ARPES can probe states above $\mathrm{E_F}$. Namely, a recent measurement has provided indirect hints of the presence of Weyl nodes in $\mathrm{Mo_{0.25}W_{0.75}Te_2}$ via a line shape analysis \cite{Belo_natCom_2016}. Following a different approach, we have measured by tr-ARPES the out-of-equilibrium dynamics of $\mathrm{MoTe_2}$, which we expect to change across the topological phase transition.  The rationale for our experiment is illustrated by Fig.\,1\,(i, j) which shows a schematic view of the bands near a type-II WP. When the VB and the CB touch and a WP appears in the WSM phase, electrons can scatter from the CB to the VB via electron -- electron scattering along the Weyl cone, as schematized in Fig.\,1\,(i). In the trivial phase the local gap between the VB and the CB is a bottleneck for the relaxation of electrons optically excited in the CB, Fig.\,1\,(j). Electrons can thermalize within the CB through electron -- electron scattering, but interband scattering towards the VB requires the exchange of energy and momentum and is mediated by phonons, with a corresponding longer relaxation time. 
%In the following, we show that the presence of efficient electron -- electron scattering channels in the WSM phase has a measurable effect on the electron relaxation time.

Figure\,2\,(a) illustrates the band dispersion along the $\mathrm{k_y=k_{WP}}$ direction measured $200$\,fs before optical excitation. The comparison of the data at low and high temperature shows no significant differences between the occupied electronic structure for the 1T' and 1T'' phases. This is confirmed by the high-resolution ARPES data \cite{SuppInfo} and by a detailed  temperature dependent study \cite{Crepaldi_PRB_17}. The effect of the optical excitation in the low-temperature 1T' phase is illustrated in Fig.\,2\,(b). It shows the difference between data taken $80$\,fs after and $200$\,fs before the arrival of the pump pulse. The color scale encodes the transfer of electrons from the occupied (blue) to the unoccupied (red) density of states, i.e. the transient electronic population in the optically excited state. 

Figure\,2\,(c-f) presents our main results. The blue and red lines show the temporal evolution of the electronic population extracted from the differential intensity map of Fig.\,2\,(b) and, respectively, from analogous data measured at $300$\,K in the 1T" phase. The signal is integrated in the small rectangular regions labeled `1' to `4'. The data show a clear shortening of the relaxation time at the lower temperature, predicted by the scenario of Fig. 1(i) as consequence of the local gap closing upon the emergence of the Weyl nodes in the 1T' phase. 

For a quantitative analysis we fitted each curve by a step function multiplied by a single decaying exponential, and broadened by a resolution-limited gaussian (green and black dashed lines).  We find that the characteristic relaxation times $\tau$ extracted from the fits are $\mathrm{\sim30-50\%}$ larger in the 1T'' topologically trivial phase (black lines) than in the 1T' WSM phase (green lines), depending on the region of the band structure where the intensity is analyzed.  Possible changes of the electronic band structure and band velocity near the WPs could affect the electron dynamics. However, our calculations show that  the only noticeable change is the emergence of WPs \cite{SuppInfo}.  Therefore we conclude that any trivial changes of the band structure have only a minor effect, and that the observed enhanced scattering rate mainly reflects the local gap closure.

%------------------------------------------------------------

The data of Fig.\,2 show that the dynamics of the excited electronic population is sensitive to the change in the unoccupied band structure. The formation of the Weyl nodes \emph{short-circuits} the local bottleneck in the dynamics of the trivial phase. There is a clear analogy between the fast dynamics at the WPs and the role played by conical intersections in the non-adiabatic relaxation of chemical systems through non-radiative transitions \cite{Polli_nature_2010}.  The bottleneck behavior associated with the formation of a local band gap has been reported in the out-of-equilibrium properties of high-temperature superconductors \cite{Giannetti_2016} and charge-density-wave systems \cite{Matthias_2016}. It is also similar to the slowdown of the dynamics in gapped bilayer graphene with respect to gapless single layer graphene \cite{Soren_PRL_14}. 

%------------------------------------------------------------

Having identified the change in electron dynamics across the topological phase transition in $\mathrm{MoTe_2}$, we now turn to the related compound $\mathrm{WTe_2}$, whose topological nature is still under debate \cite{Wu_PRB_16, Bruno_PRB_16}. Even high-resolution laser-based ARPES could not provide evidence of topological Fermi arcs \cite{Bruno_PRB_16} or of WPs, which, if present, would be located above $E_F$.
%Depending on the computational methods and lattice parameters different topological phases have been proposed \cite{Soluyanov_Nature_2015, Bruno_PRB_16, Autes_PRL_16}. 
Figure 3 summarizes our results for $\mathrm{WTe_2}$. The calculated electronic structure shows no WPs. The band structure calculated for both the experimental low-temperature ($113$\,K) \cite{Mar_1992} and high-temperature ($300$\,K) \cite{Brown_HT_structure}  lattice parameters predict the CB and VB to be always well separated in energy and momentum \cite{SuppInfo},
Fig.\,3\,(a - e). Unlike $\mathrm{MoTe_2}$, $\mathrm{WTe_2}$ has no structural phase transition and remains in the non-centrosymmetric 1T' phase.  The changes in the lattice constants between low and high temperature have only a small effect on the band dispersion, which cannot be resolved in our experiment. By comparing $\mathrm{WTe_2}$ and $\mathrm{MoTe_2}$, we find two important differences in the dynamics of $\mathrm{WTe_2}$: i) the relaxation time is temperature independent;  ii) the dynamics is much slower than in the gapped 1T'' $\mathrm{MoTe_2}$ phase. 

A quantitative analysis, summarized in Fig.\,3\,(h - k) yields characteristic times that are $\sim 3$ times larger than for 1T'' $\mathrm{MoTe_2}$. This longer relaxation time is the signature of the local gap, which acts as a bottleneck for the electron dynamics, as similarly observed for the case of 1T'' $\mathrm{MoTe_2}$. All these results are consistent with the topologically trivial nature of $\mathrm{WTe_2}$.  Notice that $\mathrm{WTe_2}$ and $\mathrm{MoTe_2}$ have very similar phonon dispersions \cite{Ma_Raman_16} and comparable Debye \cite{Callanan_JCT_92, Chen_APL_16}  and superconducting \cite{Kang_NC_15, Qi_NC_16} temperatures. Hence, electron - phonon coupling alone cannot explain the different temperature dependence measured in the two materials.
%The topologically trivial phase is also in agreement with our \emph{ab initio} calculations for the experimental lattice constants at both high and low temperature \cite{SuppInfo}
%--------------

Figure 3~(1) further illustrates the scenario that emerges from the tr-ARPES results. It sketches the relevant scattering mechanisms that determine the ultrafast dynamics in the two phases of $\mathrm{MoTe_2}$  and in $\mathrm{WTe_2}$. Blue and red indicate electron -- electron and electron -- phonon scattering processes, respectively. The former can efficiently scatter electrons from the CB to the VB only in the WSM 1T' phase of $\mathrm{MoTe_2}$. Interestingly, the calculated bands of 1T'' $\mathrm{MoTe_2}$ and $\mathrm{WTe_2}$ show that the local band gap is different in the two systems. The gap of 1T'' $\mathrm{MoTe_2}$ is indirect and the material is, also at this local level, a semimetal, so that the scattering between the CB and the VB can be mediated by phonons with no energy constraint. By contrast, $\mathrm{WTe_2}$ is locally a semiconductor, with a local direct band gap $\mathrm{E_G} = 12 (8)$ meV, for the high-(low) temperature crystal parameters. As a consequence, the lowest energy optical phonons with $\mathrm{h \nu} < 4 $ meV \cite{Ma_Raman_16} cannot contribute to the interband scattering. Hence, we attribute the longer relaxation time observed in $\mathrm{WTe_2}$ to the reduced phase space available for the electron -- electron and electron -- phonon interband scattering processes between CB and VB. This confirms the sensitivity of tr-ARPES to the small details of the unoccupied band structure. 

%-----------------------------------------------------------

In summary, tr-ARPES results bear a fingerprint of the topological transition in the relaxation dynamics of $\mathrm{MoTe_2}$. We have measured an enhancement of the relaxation rate when the local gap closes and Weyl points emerge in the low-temperature Weyl semimetal phase  of $\mathrm{MoTe_2}$. In comparison, the dynamics of $\mathrm{WTe_2}$ exhibits a slower dynamics which is compatible with a topologically trivial phase. Our work shows that the measured ultrafast dynamics is sensitive to small details of the unoccupied band structure, which are difficult to detect by other methods. Our results show that tr-ARPES can probe the consequences of topological order in novel materials.  We expect ongoing developments of theoretical microscopic models of out-of-equilibrium electron dynamics to further strengthen  the impact of this emerging technique.

%------------------------------------------------------------

We acknowledge financial support by the SNF. The research leading to these results has received funding from LASERLAB-EUROPE (grant agreement no. 654148, European Union's Horizon 2020 research and innovation programme). This work was supported in part by the Italian Ministry of University and Research under Grant Nos. FIRBRBAP045JF2 and FIRB-RBAP06AWK3 and by the European Community Research Infrastructure Action under the FP6 "Structuring the European Research Area" Program through the Integrated Infrastructure Initiative "Integrating Activity on Synchrotron and Free Electron Laser Science", Contract No. RII3-CT-2004-506008. This work has been partly performed in the framework of the nanoscience foundry and fine analysis (NFFA-MIUR Italy) project. Parts of this research were carried out at the light source PETRA III at DESY, a member of the Helmholtz Association (HGF). We would like to thank G. Hartmann, J. Buck, F. Scholz, J. Seltmann and J. Viefhaus for assistance in using beamline P04, and H. Bentmann, S. Rohlf, F. Diekmann, S. Jarausch and T. Riedel for support in using the endstation ASPHERE III.
G.A. and O.V.Y. acknowledge support by the NCCR Marvel and the ERC Starting Grant "TopoMat" (Grant No. 306504). First-principles calculations have been performed at the Swiss National Supercomputing Centre (CSCS) under Project No. s675. We thank Phil Rice for technical support. 

%------------------------------------------------------------

%_______________

%\bibliography{MoTe2_ARTEMIS_PRL}

% \bibitem[{Note(2017)}]{note} \bibinfo{journal}{Notice that the orthorhombic phase is also referred to as $\gamma$ phase \cite{Tamai_PRX_16} or T$_d$ phase \cite{Huang_NatMat_2016, Deng_NatPhy_2016, Jiang_NatCom_2016}, and that the 1T" phase is also referred to as   $\beta$ phase \cite{Tamai_PRX_16} and  also as 1T' phase \cite{Huang_NatMat_2016, Deng_NatPhy_2016, Jiang_NatCom_2016}.}

%_______________

%_______________

\end{document}